\documentclass[twoside,twocolumn,9pt]{article}
\usepackage{extsizes}
\usepackage[super,sort&compress,comma]{natbib} 
\usepackage[version=3]{mhchem}
\usepackage[left=1.5cm, right=1.5cm, top=1.785cm, bottom=2.0cm]{geometry}
\usepackage{amsmath}
\usepackage{amsfonts}
\usepackage{amssymb}
\usepackage{amsbsy}
\usepackage{balance}
\usepackage{mathptmx}
\usepackage{sectsty}
\usepackage{graphicx} 
\usepackage{lastpage}
\usepackage[format=plain,justification=justified,singlelinecheck=false,font={stretch=1.125,small,sf},labelfont=bf,labelsep=space]{caption}
\usepackage{float}
\usepackage{fancyhdr}
\usepackage{fnpos}
\usepackage[english]{babel}
\addto{\captionsenglish}{%
  
}
\usepackage{array}
\usepackage{droidsans}
\usepackage{charter}
\usepackage[T1]{fontenc}
\usepackage[usenames,dvipsnames]{xcolor}
\usepackage{setspace}
\usepackage[compact]{titlesec}
\usepackage{hyperref}
\newcommand{\lla}{\left\langle}
\newcommand{\rra}{\right\rangle}

\definecolor{cream}{RGB}{222,217,201}

\begin{document}

\pagestyle{fancy}
\thispagestyle{plain}
\fancypagestyle{plain}{
%%%HEADER%%%
\renewcommand{\headrulewidth}{0pt}
}
%%%END OF HEADER%%%

%%%PAGE SETUP - Please do not change any commands within this section%%%
\makeFNbottom
\makeatletter
\renewcommand\LARGE{\@setfontsize\LARGE{15pt}{17}}
\renewcommand\Large{\@setfontsize\Large{12pt}{14}}
\renewcommand\large{\@setfontsize\large{10pt}{12}}
\renewcommand\footnotesize{\@setfontsize\footnotesize{7pt}{10}}
\makeatother

\renewcommand{\thefootnote}{\fnsymbol{footnote}}
\renewcommand\footnoterule{\vspace*{1pt}% 
\color{cream}\hrule width 3.5in height 0.4pt \color{black}\vspace*{5pt}} 
\setcounter{secnumdepth}{5}

\makeatletter 
\renewcommand\@biblabel[1]{#1}            
\renewcommand\@makefntext[1]% 
{\noindent\makebox[0pt][r]{\@thefnmark\,}#1}
\makeatother 
\renewcommand{\figurename}{\small{Fig.}~}
\sectionfont{\sffamily\Large}
\subsectionfont{\normalsize}
\subsubsectionfont{\bf}
\setstretch{1.125} %In particular, please do not alter this line.
\setlength{\skip\footins}{0.8cm}
\setlength{\footnotesep}{0.25cm}
\setlength{\jot}{10pt}
\titlespacing*{\section}{0pt}{4pt}{4pt}
\titlespacing*{\subsection}{0pt}{15pt}{1pt}
%%%END OF PAGE SETUP%%%

%%%FOOTER%%%
\fancyfoot{}
\fancyfoot[LO,RE]{\vspace{-7.1pt}\includegraphics[height=9pt]{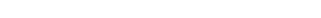}}
\fancyfoot[CO]{\vspace{-7.1pt}\hspace{13.2cm}\includegraphics{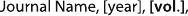}}
\fancyfoot[CE]{\vspace{-7.2pt}\hspace{-14.2cm}\includegraphics{RF}}
\fancyfoot[RO]{\footnotesize{\sffamily{1--\pageref{LastPage} ~\textbar  \hspace{2pt}\thepage}}}
\fancyfoot[LE]{\footnotesize{\sffamily{\thepage~\textbar\hspace{3.45cm} 1--\pageref{LastPage}}}}
\fancyhead{}
\renewcommand{\headrulewidth}{0pt} 
\renewcommand{\footrulewidth}{0pt}
\setlength{\arrayrulewidth}{1pt}
\setlength{\columnsep}{6.5mm}
\setlength\bibsep{1pt}
%%%END OF FOOTER%%%

%%%FIGURE SETUP - please do not change any commands within this section%%%
\makeatletter 
\newlength{\figrulesep} 
\setlength{\figrulesep}{0.5\textfloatsep} 

\newcommand{\topfigrule}{\vspace*{-1pt}% 
\noindent{\color{cream}\rule[-\figrulesep]{\columnwidth}{1.5pt}} }

\newcommand{\botfigrule}{\vspace*{-2pt}% 
\noindent{\color{cream}\rule[\figrulesep]{\columnwidth}{1.5pt}} }

\newcommand{\dblfigrule}{\vspace*{-1pt}% 
\noindent{\color{cream}\rule[-\figrulesep]{\textwidth}{1.5pt}} }

\makeatother
%%%END OF FIGURE SETUP%%%

%%%TITLE, AUTHORS AND ABSTRACT%%%
\twocolumn[
  \begin{@twocolumnfalse}
{\includegraphics[height=30pt]{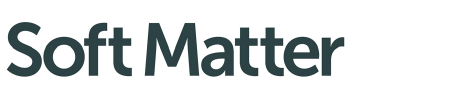}\hfill\raisebox{0pt}[0pt][0pt]{\includegraphics[height=55pt]{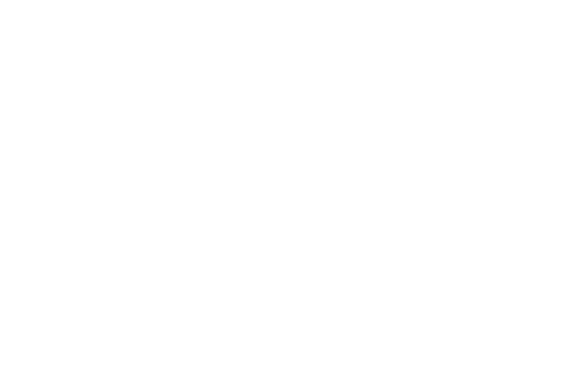}}\\[1ex]
\includegraphics[width=18.5cm]{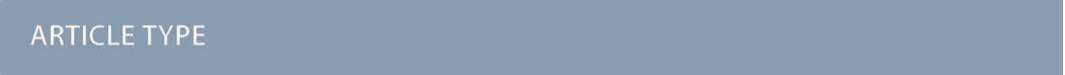}}\par
\vspace{1em}
\sffamily
\begin{tabular}{m{4.5cm} p{13.5cm} }

\includegraphics{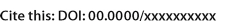} & \noindent\LARGE{\textbf{Dynamical behavior of compound vesicles in wall-bounded shear flow}} \\%Article title goes here instead of the text "This is the title"
\vspace{0.3cm} & \vspace{0.3cm} \\

 & \noindent\large{Antonio Lamura\textit{$^{a}$}}\\%Author names go here instead of "Full name", etc.

\includegraphics{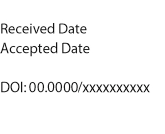} & \noindent\normalsize{We report a numerical study addressing
the dynamics of compound vesicles confined in a channel under shear flow. The system comprises
a smaller vesicle embedded within a larger one and can be used to mimic, for example, leukocytes or
nucleate cells. A two-dimensional model, which combines molecular dynamics and mesoscopic hydrodynamics
including thermal fluctuations, is adopted to perform an extended investigation. 
We are able to vary independently the swelling degree and the relative size of vesicles, the viscosities
of fluids internal and external to vesicles, and the Capillary number, so to observe a rich dynamical phenomenology
which goes well beyond what observed for single vesicles, matching quantitatively with experimental findings.
Tank-treading, tumbling, and trembling motions are enriched by dynamical states where inner and outer vesicles
can perform different motions. We show that thermal fluctuations are crucial during trembling and swinging dynamics,
as observed in experiments. Undulating motion of the external vesicle, characterized by periodic oscillation
of the inclination and buckling of the membrane, is observed at high filling fractions. 
This latter state exhibits features that are shown to depend on the relative size, the swelling degree of both
vesicles as well as on thermal noise lacking in previous analytical and numerical studies.
} 

\end{tabular}

 \end{@twocolumnfalse} \vspace{0.6cm}

  ]
%%%END OF TITLE, AUTHORS AND ABSTRACT%%%

%%%FONT SETUP - please do not change any commands within this section
\renewcommand*\rmdefault{bch}\normalfont\upshape
\rmfamily
\section*{}
\vspace{-1cm}

%%%FOOTNOTES%%%

\footnotetext{\textit{$^{a}$~Istituto Applicazioni Calcolo, Consiglio Nazionale delle Ricerche (CNR), Via Amendola 122/D, 70126 Bari, Italy; E-mail: antonio.lamura@cnr.it}}

%%%MAIN TEXT%%%%
\section{Introduction}

Compound vesicles, which comprise a smaller vesicle embedded
within a larger one, are systems of relevant biological importance
since can be seen as biomimetic models for
 multi-compartmentalized
cells such as leukocytes \cite{schm80,veer11,kaou13,pan2015,sinh19} or nucleate cells
\cite{haro13,pana14,kuma24}. 
Furthermore, vesicles embedding one or even more vesicles, as in the case
of vesosomes \cite{walk97,kisa04}, are also of interest for drug 
delivery applications
\cite{giul21},
where controlled deformation and compartmentalization can be
properly exploited \cite{saad23}. 
Compound vesicles under shear flow have attracted attention due to
their relevance in physiological conditions and microfluidic systems, 
where cells or synthetic vesicles experience different
dynamical states \cite{leva14}. The study of their dynamical behavior requires
to take into account inter-membrane interactions,
membrane dynamics, and fluid-structure coupling when exposed to external flow. 

The dynamical responses of single vesicles in shear flow, including
tank-treading (TT), tumbling (TU), and trembling (TR) motions, 
have been extensively studied theoretically \cite{kell82,misb06,lebe07,dank07b,fink08}, 
numerically \cite{nogu04,nogu05,nogu07bis,mess09,kaou11,zhao11,nait18,lamu22},  and experimentally \cite{kant05,kant06,vitk08,zabu11,afik16}.
However, the introduction of an internal vesicle produces a richer
dynamics since extra degrees
of freedom are added such as internal vesicle rotation, displacement, 
and deformation, 
as well as hydrodynamic coupling between the internal and external membrane.
These features make compound
vesicles a system suitable to understand the interplay between membranes
dynamics under geometrical constraints and hydrodynamic interactions.

Recent studies have revealed, in addition to motions observed for
single vesicles,
novel dynamical modes unique to compound vesicles, such as synchronized or
asynchronous TU of the internal and external vesicle \cite{leva14,sinh19},
swinging (SW) of the internal vesicle, whose main axis oscillates around
a positive mean angle, 
at large shear rates \cite{veer11,leva14}, 
and undulating (UND) motion where 
the external vesicle oscillates around a positive mean angle while its
shape undergoes large undulations \cite{kaou13}.
Nevertheless, the parameter space governing these
behaviors - 
including reduced volumes of internal and external vesicles,
ratio of the viscosity of the fluid around the internal vesicle
to the viscosity of the outer fluid,
Capillary number, and occupied fraction of the interior volume -
remains only partially mapped, and many questions persist
about dynamical behaviors. 
Indeed, theoretical approaches \cite{veer11,sinh19}
are limited to consider vesicles in the quasi-spherical approximation,
available numerical simulations \cite{veer11,kaou13} neglect thermal fluctuations 
that are known to be relevant in TR \cite{leva12,abre13,abre14,leva14},
and experimentally it is not trivial to control physical and geometrical
parameters of compound vesicles \cite{leva14}.

In the present study, we aim at exploring the behavior of
compound vesicles through numerical simulations
accessing wide ranges of the parameters controlling the system. 
Our results are achieved by using
mesoscale hydrodynamic simulations of a two-dimensional model system, 
including thermal fluctuations.
These are expected \cite{sinh19} to be crucial in order to capture features observed only in experiments \cite{leva14}.
We can vary 
the swelling degree of both vesicles going from deflated
to quasi-circular shapes, to tune independently
 viscosities for the fluids around
and inside each vesicle, 
to change the Capillary number,
and to consider different sizes of the internal vesicle.
In this way, we are able to observe a rich phenomenology in the dynamical 
behavior of compound vesicles which well matches with experimental observations
of Ref.~\cite{leva14}. 
Moreover, our study allows a full characterization of the
UND motion under different conditions to an extent never observed before.
 
\section{Numerical model} \label{sec:model}

\begin{figure}[h]
\begin{center}
\includegraphics*[width=.48\textwidth,angle=0]{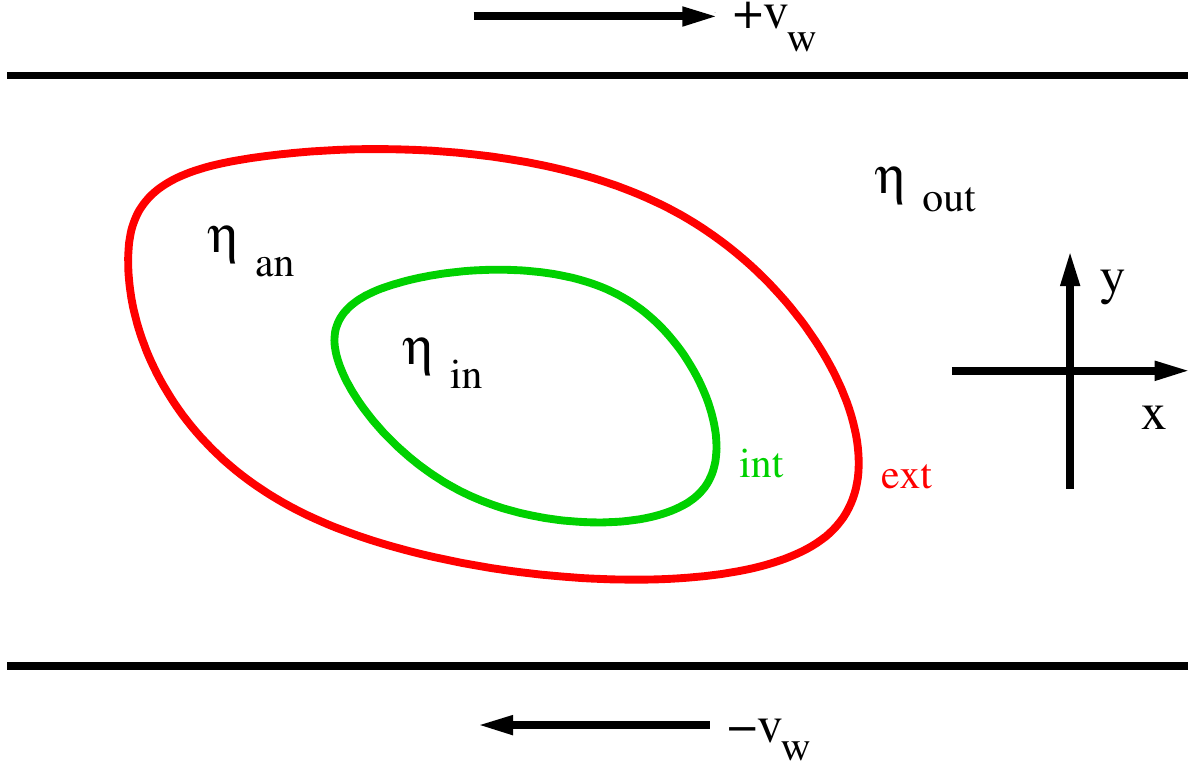}
\caption{Schematic layout of the simulated system.
\label{fig:layout}
}
\end{center}
\end{figure}

We consider a two-dimensional system  
consisting of a compound vesicle, made of one internal ({\it int})
vesicle embedded in an external ({\it ext})
one, which is suspended in 
a shearing fluid (see Fig.~\ref{fig:layout}).
The fluid in each of the three
zones, separated by the vesicle membranes,
is described by the multi-particle collision dynamics 
\cite{male99,male00,kapr08,gomp09,howa19} with viscosities
$\eta_{out}, \eta_{an}, \eta_{in}$ for the outer ({\it out}), annular ({\it an}),
and inner ({\it in}) areas, respectively. The system of size 
$L_x \times L_y$ is placed between two horizontal walls 
sliding along the $x$-direction (flow direction) with velocities
$\pm v_w$. 
Periodic boundary conditions are imposed along the flow direction
and bounce-back boundary conditions are implemented at the walls
\cite{lamu01,lamu02}. This allows us to simulate
a linear flow profile $(\dot\gamma y,0)$
with shear rate $\dot\gamma= 2 v_w /L_y$, 
$y$ being the vertical coordinate along the shear direction.
Each vesicle is modeled by enforcing the conservation of
internal area $A_0$ and perimeter $L_0$ of the enclosing membrane,
and simulated via molecular dynamics.
Internal and external membranes are
characterized by the same bending rigidity $\kappa$
and experience a mutual repulsion at short distances \cite{lamu13}.
Solvent particles and membranes interact via three-body collisions which conserve
linear and angular momenta \cite{mess09}.
The details for the numerical implementation of
fluid, vesicles, and solvent-membrane coupling can be found in 
Refs.~\cite{lamu15,lamu22}.
It is useful for the following analysis
to introduce
the excess length
$\Delta=L_0/\sqrt{A_0/\pi}- 2 \pi$.
$\Delta$ is a measure of the swelling degree, being  $\Delta > 0$ 
in the case of deflated vesicles and $\Delta=0$ for circular ones.
The vesicle dynamical states in shear flow can be characterized by the
dimensionless quantities
$\Lambda=(32+23 \lambda) \sqrt{\Delta_{ext}}/(8 \sqrt{30 \pi})$ and
$S=14 \pi \dot\gamma^* /(3 \sqrt{3} \Delta_{ext})$ \cite{lebe07}
where $\lambda=\eta_{an}/\eta_{out}$ is the viscosity contrast
and $\dot\gamma^*=\dot\gamma \eta_{out} R_{0,ext}^3/\kappa$ is the
Capillary number (or reduced shear rate), 
$R_0=L_0 / (2 \pi)$ being the vesicle radius.
The use of $\Lambda$ and $S$
will allow us to make easy comparison with the experimental
results \cite{leva14}. 
The compound vesicle dynamics is also affected by
the occupied fraction of the interior area of the external vesicle,
defined to be $\phi=R_{0,int}/R_{0,ext}$ \cite{kaou13,leva14,sinh19},
and by the excess length $\Delta_{int}$ of the internal vesicle \cite{leva14}.

The novelty of the present study is that we are able
to vary all the four parameters $\Lambda, S, \phi, \Delta_{int}$
in wide ranges,
accessing different dynamical states of the compound vesicle.
In the following we set
$0.16 \leq \Delta_{ext} \leq 1.23$, $0.16 \leq \Delta_{int} \leq 0.74$,
$0.2 \leq \phi \leq 0.8$, $1 \leq \lambda \leq 15$, 
$1 \leq \dot\gamma^* \leq 20$, $\eta_{in}=\eta_{out}$,
 $L_x/R_{0,ext}=12.04$, and $L_y/R_{0,ext}=5.76$, while maintaining
fixed $R_{0,ext}$.
The bending rigidity is chosen to be $\kappa=3.27 k_B T R_{0,ext}$,
where $k_B T$ is the thermal energy of the system.
This value of $\kappa$ gives rise to a comparable
amplitude of undulation modes as 
for compound vesicles in experiments \cite{leva14}
(see the following Fig.~\ref{fig:tt} (c)).
All simulations are run by keeping the
inertial effects negligible: This is assured
by making the Reynolds number
$Re=\dot\gamma \rho_{out} R_{0,ext}^2/\eta_{out} \lesssim 0.1$, $\rho_{out}$
being the outer fluid density.

\section{Results} \label{sec:results}

 In this Section we show and characterize the different dynamical states of compound vesicles
discussing the results and comparing them with previous studies.

\subsection{ Tank-treading motion}

At values $\Lambda \lesssim 1$ and $10 \lesssim S \lesssim 10^3$, compound vesicles are found in a steady TT regime. Both 
internal and external vesicle keep a steady inclination angle $\theta$, defined as the
angle between the vesicle main long axis and the flow direction,
while the membranes perform
a tank-treading motion with a characteristic frequency $\omega$. 
\begin{figure}[h]
\begin{center}
\includegraphics*[width=0.48\textwidth,angle=0]{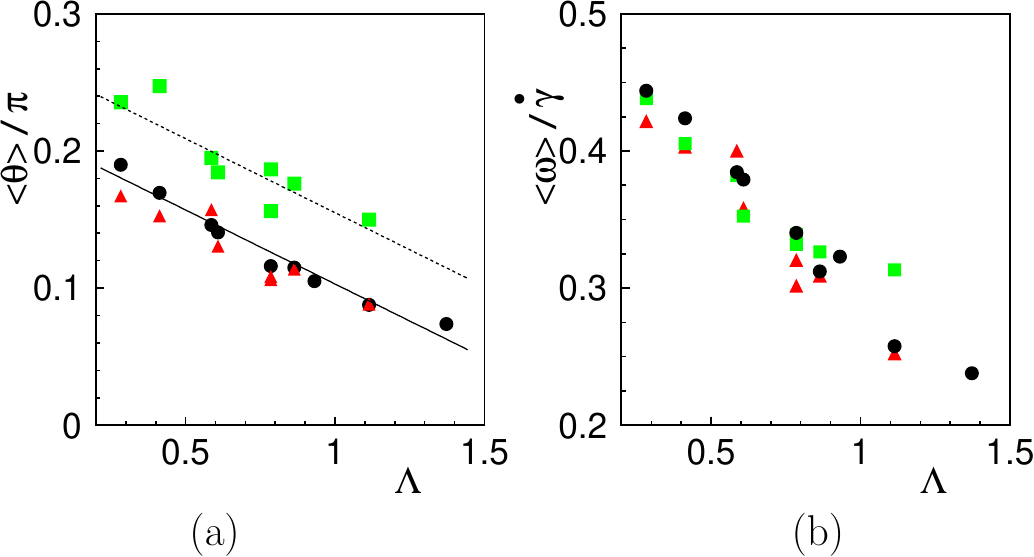}
\caption{(a) Dependence of the average inclination angle on $\Lambda$
for the external (red triangles) and internal (green squares) tank-treading
vesicles
compared with data of single tank-treading vesicles (black circles)
\cite{lamu22}.
The full line is the fit to the data of single vesicles and the dashed line
is the same fit line shifted up. The data shown are obtained from different
compound vesicles with $0.16 \leq \Delta_{ext} \leq 1.23$,
$0.16 \leq \Delta_{int} \leq 0.74$,
and $0.4 \leq \phi \leq 0.8$. In the case of single vesicles it is
$0.16 \leq \Delta \leq 1.23$.
(b)  Dependence of the average tank-treading frequency on $\Lambda$
for the external (red triangles) and internal (green squares) vesicles
compared with data of single vesicles (black circles)
\label{fig:tt}
}
\end{center}
\end{figure}
\begin{figure}[h]
\begin{center}
\includegraphics*[width=0.48\textwidth,angle=0]{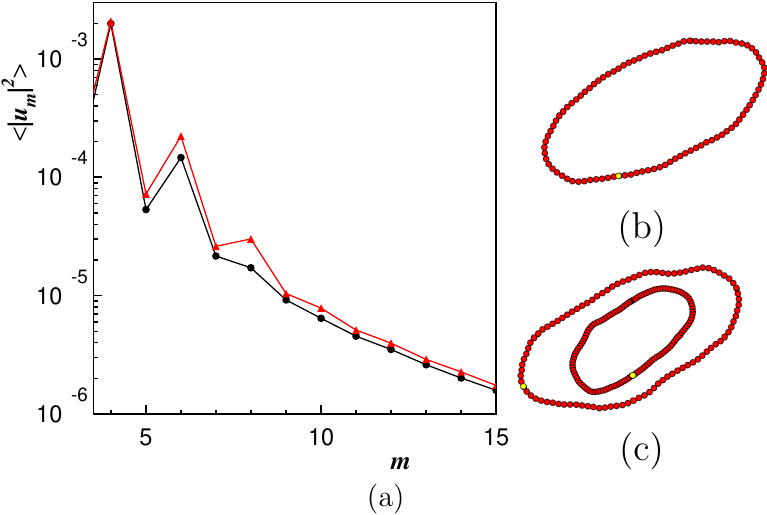}
\caption{(a) Comparison of the spectra, averaged in time, of amplitudes of 
shapes for a single vesicle with $\Delta=0.74$ (black circles) and
a compound one with $\Delta_{ext}=\Delta_{int}=0.74$,
$\phi=0.6$ (red triangles).
Both single vesicle and compound one are in tank-treading motion at $(\Lambda, S)=(0.61, 229)$.
Corresponding typical configurations are shown in panels (b) and (c), respectively.
\label{fig:modes}
}
\end{center}
\end{figure}
This motion is observed for different swelling degrees of internal and external vesicles in a wide
range of the filling fraction ($0.4 \leq \phi \leq 0.8$).
In order to compute the inclination angle, 
the two eigenvalues $\lambda_M$ and $\lambda_m$, 
with $\lambda_M > \lambda_m$, of the vesicle gyration tensor are extracted. 
The inclination angle is defined as the angle formed by the eigenvector corresponding
to $\lambda_M$ with the flow direction.
The average inclination angles $\lla \theta \rra$ of compound vesicles are shown in Fig.~\ref{fig:tt} (a). Data of the external vesicle (red triangles) show a monotonic decrease with
$\Lambda$. Moreover,  the values of $\lla \theta \rra_{ext}$
are similar to the ones observed for single 
vesicles \cite{lamu22}, which are reported in the same figure (black circles) and 
can be quite well fitted by a straight line.
The comparable values of the inclination angles suggest 
that the transition at $\lla \theta \rra \simeq 0$ from the TT regime to the TR one
occurs at similar value $\Lambda_c \simeq 1.91$, and that
the internal vesicle has negligible effect on the external dynamics. 
On the other hand, the internal vesicle attains larger inclination angles (green squares) with the same dependence on 
$\Lambda$ so that the data can be interpolated by simply shifting up the fit line 
when using the external flow parameters $\Lambda$ and $S$.
The larger values of $\lla \theta \rra_{int}$ might be due to the confinement exerted by the
external vesicle, a phenomenon already observed for single vesicles \cite{kaou12}.
As a consequence, the relative angle $\lla \theta \rra_{int} - \lla \theta \rra_{ext}$
does not depend on $\Lambda$ and $\phi$.
These results are fully consistent with the ones observed in the only available
experiments on sheared compound vesicles \cite{leva14}. 
Numerical simulations of a compound vesicle with a solid spherical inclusion \cite{veer11}
and with an inner vesicle \cite{kaou13,sinh19} show a dependence of $\lla \theta \rra_{ext}$
and $\lla \theta \rra_{int}$ on $\phi$. Moreover, differences in
the values of  $\lla \theta \rra_{ext}$ and $\Lambda_c$,
with respect to the case of a single vesicle, are observed. These discrepancies can be attributed to the lack
of thermal noise, which is known to affect the TT-TR transition \cite{abre14},
in previous simulations.
We can also compute the tank-treading frequencies $\omega$ whose
average values, rescaled by the shear rate $\dot\gamma$, 
are shown for the external
(red triangles) and internal (green squares) membranes of a compound vesicle and for a single
vesicle (black circles) in Fig.~\ref{fig:tt} (b). Data points refer to the same systems
presented in the panel (a).
Remarkably, it appears that the internal and external membranes as well as the membrane of single vesicle
rotate with comparable rescaled average frequencies which show the same dependence on $\Lambda$,
while being
not dependent on both the filling fractions and the excess lengths. 

\begin{figure}[h!]
\begin{center}
\includegraphics*[width=.48\textwidth,angle=0]{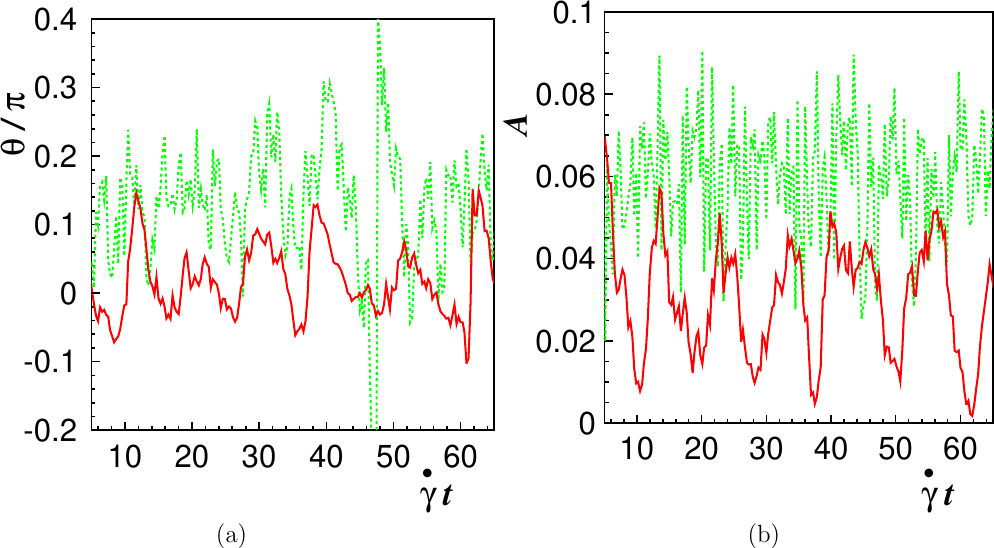}
\caption{(a) Time evolution of the inclination angles
$\theta_{ext}$ (red continuous line)
and $\theta_{int}$ (green dashed line) of a compound vesicle with
$\Delta_{ext}=\Delta_{int}=0.16$, $\phi=0.2$ at $(\Lambda, S)=(1.94, 1058)$
showing trembling (external vesicle) and swinging (internal vesicle) motion.
(b) Time evolution of the asphericities $A_{ext}$ (red continuous line)
and $A_{int}$ (green dashed line) corresponding
to the case in panel (a).
\label{fig:trsw}
}
\end{center}
\end{figure}
In order to quantitatively characterize the shapes of the external vesicle during the TT motion, we parametrize the membrane contour as a function of the polar angle $\varphi$ ($0 \leq \varphi \leq 2 \pi$)
by writing the position
$\mathbf{r}(\varphi)$ of any point on the membrane as
%\begin{equation}
    $\mathbf{r}(\varphi)=R_{0,ext} \mathbf{e}_r(\varphi)\Big [ 1+ u(\varphi) \Big ]$,
%\end{equation}
where $\mathbf{e}_r(\varphi)$ is a unit vector along the radial direction and $u(\varphi)$
is the dimensionless shape deformation. Since $u(\varphi)$ is a real periodic function
of $\varphi$, it can expanded via Fourier transform as
%\begin{equation}
    $u(\varphi) = \sum_m u_m \exp{(i m \varphi)} $.
%\end{equation}
The computed stationary correlations of the complex Fourier modes $u_m$ are shown in Fig.~\ref{fig:modes} (a)
for a compound vesicle with $\Delta_{ext}=\Delta_{int}=0.74$ and
$\phi=0.6$, and for
 a single vesicle with the same excess length.
Both single vesicle and compound one are in tank-treading motion with flow
parameters $(\Lambda, S)=(0.61, 229)$. The main feature that appears, is the amplification of the
even modes of the external vesicle spectrum at $m \geq 6$ (red triangles) compared
with those measured for the single vesicle (black circles). This is due to the fact that the 
shape of the external vesicle is modified in the compound one, being more parallelogram-like
(see Fig.~\ref{fig:modes} (c)),
with respect to the ellipsoidal shape of the single vesicle (see Fig.~\ref{fig:modes} (b)) \cite{fink08,lamu22}. 
These features match with what observed in experiments \cite{leva14}
where the measured values of $\lla |u_m|^2 \rra$
 (see Fig.~2 of Ref.\cite{leva14}) are in a good quantitative agreement with the present ones.

\begin{figure*}[h!]
\begin{center}
\includegraphics*[width=0.8\textwidth,angle=0]{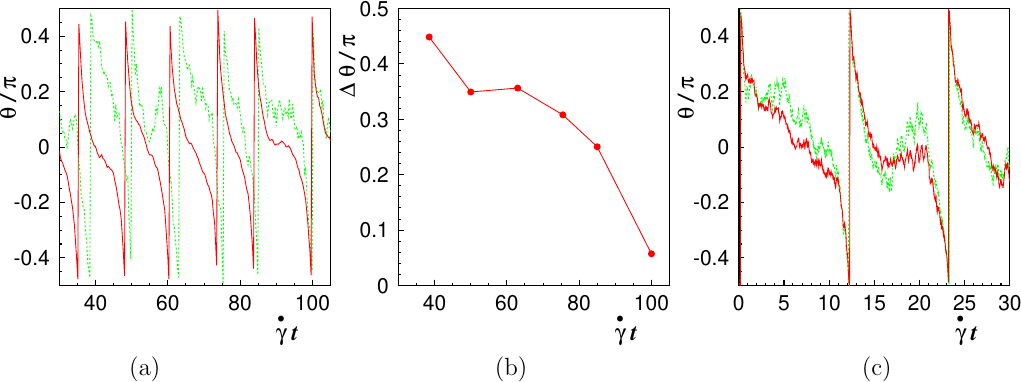}
\caption{
(a)  Time evolution of the inclination
 angles $\theta_{ext}$ (red continuous line)
and $\theta_{int}$ (green dashed line) of a compound vesicle with
$\Delta_{ext}=1.23$, $\Delta_{int}=0.74$,
$\phi=0.2$ at $(\Lambda, S)=(5.38, 138)$
in desynchronized tumbling motion.
(b) Time dependence of $\Delta \theta=\theta_{int}-\theta_{ext}$ corresponding
to the data in panel (a).
(c) Time evolution of the inclination angles
$\theta_{ext}$ (red continuous line)
and $\theta_{int}$ (green dashed line) of a compound vesicle with
$\Delta_{ext}=\Delta_{int}=0.74$,
$\phi=0.6$ at $(\Lambda, S)=(4.18, 23)$ in synchronized
tumbling motion.
\label{fig:tilt_sincasincr}
}
\end{center}
\end{figure*}

\subsection{ Trembling motion}

Upon increasing $\Lambda$ beyond $\Lambda_c$, 
the external vesicle performs TR motion during which its main axis oscillates around the
flow direction while undergoing strong shape modifications, as found for single
vesicles \cite{kant07,leva14_bis}. The inner vesicle is observed to perform 
SW motion, which is  characterized by large variations of the inclination angle around a positive mean value, 
as depicted in Fig.~\ref{fig:trsw} (a) for small $\phi$ at $(\Lambda,S)=(1.94,1058)$. 
It can be seen that at $\dot\gamma \simeq 48$ an isolated tumbling event occurs. This
reminds the intermediate regime between TU and SW observed in
simulations of quasi-spherical compound vesicles \cite{sinh19}.
In order to gain information about the
shape deformation, the asphericity 
$A=\Big [ ( \lambda_M - \lambda_m ) / (\lambda_M + \lambda_m ) \Big ]^2$ is computed,
being $A=0$ for circular shapes.
The time evolution of $A$ for the internal and external vesicles is shown in
Fig.~\ref{fig:trsw} (b) where it can be appreciated that $A_{ext}$ widely oscillates in time, appearing synchronized with $\theta_{ext}$, as in the case of single vesicles \cite{zhao11}. 
The supplementary animation file movie1.mp4 provides a comprehensive view of the TR/SW motion.
A similar dynamics has been observed for comparable values of $\phi$, $\Lambda$, and $S$
in the experiments of Ref.~\cite{leva14} (see Fig.~4 therein).
The SW motion has been also seen for the external vesicle of a vesicle embedding
a solid ellipsoidal inclusion which performs TU motion \cite{veer11}.
The internal vesicle has been found in SW state also in previous simulations of compound vesicles \cite{kaou13,sinh19}.

\begin{figure}[h!]
\begin{center}
\includegraphics*[width=.48\textwidth,angle=0]{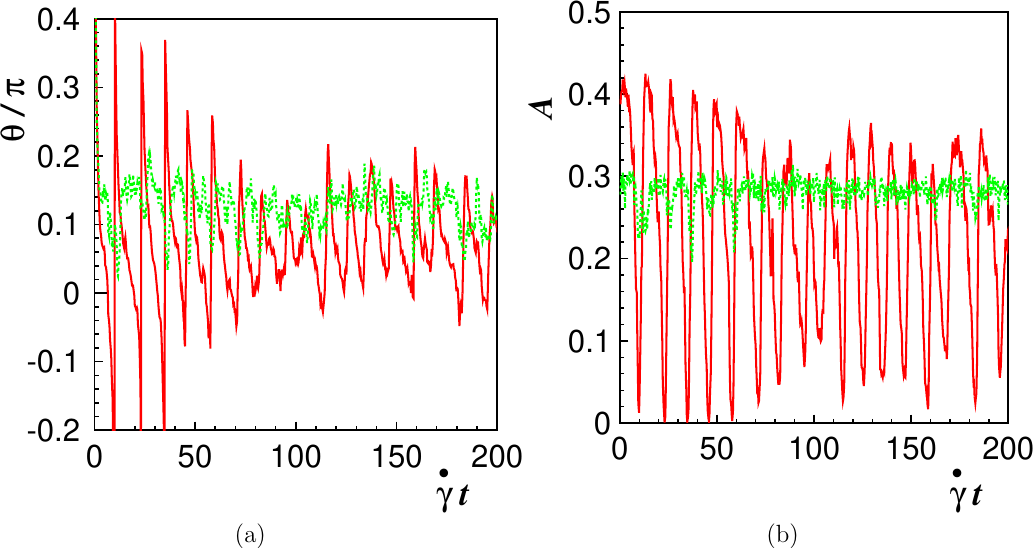}
\caption{(a) Time evolution of the inclination angles
$\theta_{ext}$ (red continuous line)
and $\theta_{int}$ (green dashed line) of a compound vesicle with
$\Delta_{ext}=1.23$, $\Delta_{int}=0.74$,
$\phi=0.6$ at $(\Lambda, S)=(5.38, 138)$
showing undulating (external vesicle) and swinging (internal vesicle) motion.
(b) Time evolution of the asphericities $A_{ext}$ (red continuous line)
and $A_{int}$ (green dashed line) corresponding
to the case in panel (a).
\label{fig:und_lambda15_phi06_gam20}
}
\end{center}
\end{figure}

\subsection{ Tumbling motion}

A further increase of $\Lambda$ determines the transition of the compound vesicle
to TU motion: both internal
and external vesicle rotate as solid-like elongated particles. Their angular velocities
are triggered by the fraction $\phi$ of the occupied interior area.
At small $\phi$, the internal and external vesicle motions are weakly correlated.
Their centers are not stationary, differently from what happens in studies where thermal
noise is discarded \cite{kaou13,sinh19}, and their TU velocities slightly deviate from each other.
This desynchronizes $\theta_{int}(t)$ and $\theta_{ext}(t)$, as shown
in Fig.~\ref{fig:tilt_sincasincr} (a) for $\phi=0.2$ at $(\Lambda,S)=(5.38,138)$. The relative difference between the inclination angles
is depicted in panel (b) of Fig.~\ref{fig:tilt_sincasincr} in order to quantify the aforementioned
desynchronization. Larger values of $\phi$ allow the internal and external vesicles to tumble
in a synchronous way, as illustrated in Fig.~\ref{fig:tilt_sincasincr} (c) at $(\Lambda,S)=(4.18,23)$. These features
are in common with the theoretical analysis
of Ref.~\cite{sinh19} for quasi-spherical compound vesicles,
 and have been as well observed in the experiments of Ref.~\cite{leva14}
(see Figs.~5-6 therein) for comparable values of the control parameters $\Lambda, S, \phi, \Delta_{int}$.

\begin{figure}[h!]
\begin{center}
\includegraphics*[width=.48\textwidth,angle=0]{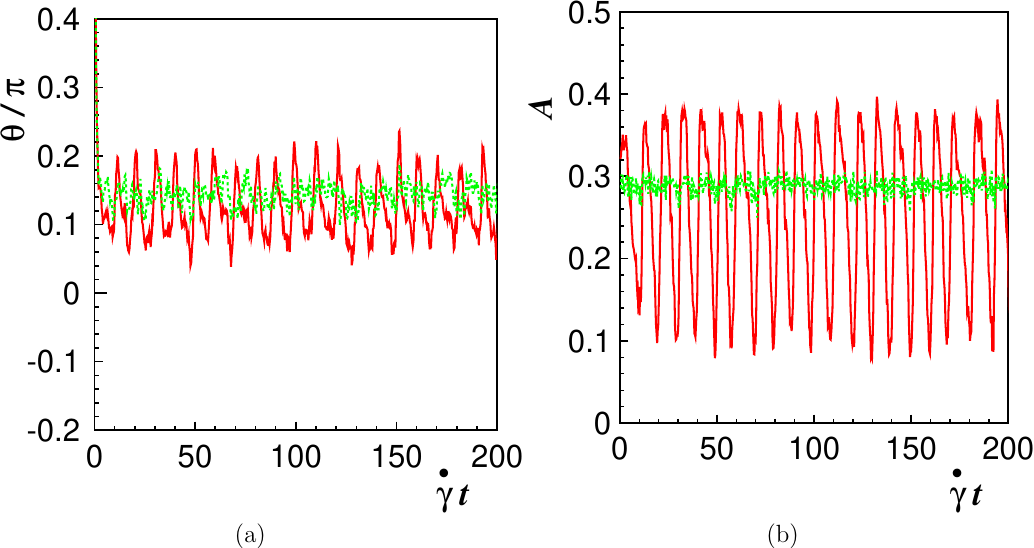}
\caption{(a) Time evolution of the inclination angles
$\theta_{ext}$ (red continuous line)
and $\theta_{int}$ (green dashed line) of a compound vesicle with
$\Delta_{ext}=1.23$, $\Delta_{int}=0.74$,
$\phi=0.8$ at $(\Lambda, S)=(5.38, 138)$
showing undulating (external vesicle) and tank-treading
(internal vesicle) motion.
(b) Time evolution of the asphericities $A_{ext}$ (red continuous line
and $A_{int}$ (green dashed line) corresponding
to the case in panel (a).
\label{fig:und_lambda15_phi08_gam20}
}
\end{center}
\end{figure}
\begin{figure}[h!]
\begin{center}
\includegraphics*[width=.4\textwidth,angle=0]{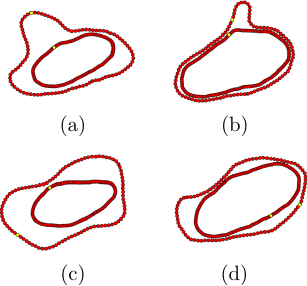}
\caption{Configurations of compound vesicles 
with
$\Delta_{ext}=1.23$, $\Delta_{int}=0.74$,
$\phi=0.6$ (a), $0.8$ (b) at $(\Lambda, S)=(5.38, 138)$
and with
$\Delta_{ext}=\Delta_{int}=0.74$,
$\phi=0.6$ (c), $0.8$ (d) at $(\Lambda, S)=(4.18, 229)$.
\label{fig:conf_und}
}
\end{center}
\end{figure}

\subsection{ Undulating motion}

For large values
of $\Lambda$ and $S$ and high filling fraction ($\phi \gtrsim 0.6$), the UND motion of the external vesicle appears. This is characterized by regular oscillations of the inclination angle $\theta_{ext}(t)$ around a positive mean angle $\lla \theta_{ext} \rra$
with wide shape undulations \cite{kaou13} during the rotation of the membrane. 
By increasing the filling fraction $\phi$, the amplitudes of 
$\theta_{ext}(t)$ reduce while  the average value $\lla \theta_{ext} \rra$ slightly increases.
$\phi$ affects the motion of the internal vesicle, as well, 
that can perform either SW, as shown in Fig.~\ref{fig:und_lambda15_phi06_gam20} (a) for $\phi=0.6$
(see also the supplementary animation file movie2.mp4), or
TT when increasing $\phi$, as shown in Fig.~\ref{fig:und_lambda15_phi08_gam20} (a) 
for $\phi=0.8$
(see also the supplementary animation file movie3.mp4). In both cases, the flow parameters are $(\Lambda, S)=(5.38, 138)$ and the external vesicle is more deflated than  the internal one, being $\Delta_{ext}=1.23$ and $\Delta_{int}=0.74$.
It is interesting to note that during the UND motion, the external membrane buckles assuming lobated shapes which depend on the filling fraction,
as shown in Fig.~\ref{fig:conf_und} (a), (b). 
\begin{figure}[h!]
\begin{center}
\includegraphics*[width=.48\textwidth,angle=0]{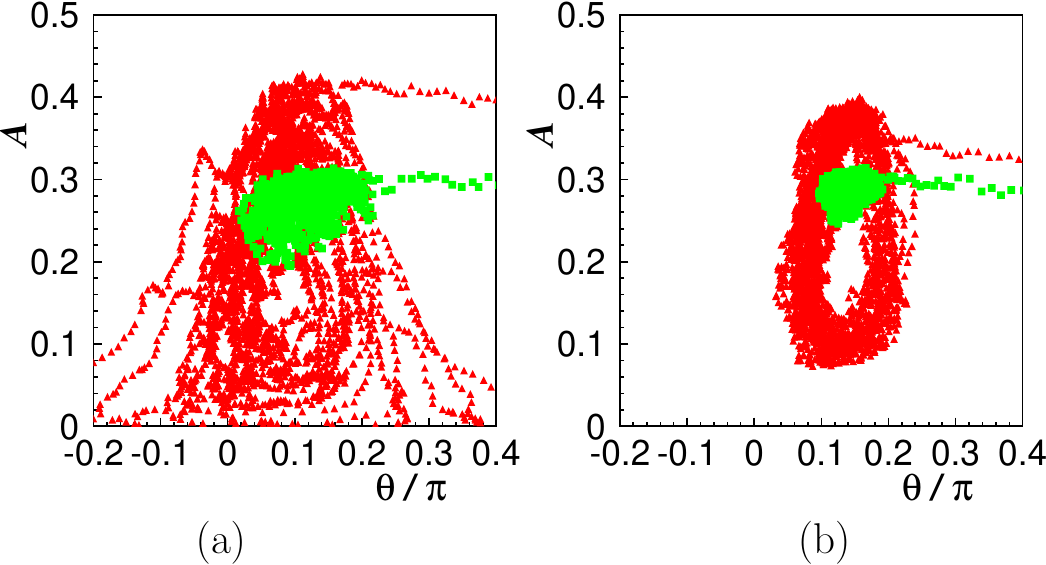}
\caption{Asphericity $A$ as a function of the inclination angle $\theta$
for the external (red triangles) and internal (green squares) vesicles
corresponding in panels (a) and (b) to the cases shown in Fig.~\ref{fig:und_lambda15_phi06_gam20} 
and Fig.~\ref{fig:und_lambda15_phi08_gam20}, respectively.
\label{fig:asphertilt_und}
}
\end{center}
\end{figure}
The shape variations are quantified in the panels (b) of Figs.~\ref{fig:und_lambda15_phi06_gam20} and \ref{fig:und_lambda15_phi08_gam20}
where the time behaviors of the asphericities $A_{ext}$ and $A_{int}$ are depicted. 
It is evident
that $A_{ext}$ oscillates with the same frequency of $\theta_{ext}$, reaching a local maximum when the inclination angle is at its maximum.
It has to be remarked that the amplitudes of $A_{ext}(t)$ during UND motion
are much larger
compared to those ones observed in TR motion (see Fig.~\ref{fig:trsw} (b)). Moreover,
the oscillatory pattern can be used to determine a characteristic frequency $\omega_A$
by performing a spectral analysis of the time series $A_{ext}(t)$.
Interestingly, it is found that $\omega_A \simeq 2 \lla \omega \rra_{ext}$
where $\lla \omega \rra_{ext}$ is the average rotation frequency of the external vesicle.
This result corresponds to the fact that the external vesicle experiences stretching along
the $\pi/4$ direction and compression along the $-\pi/4$ direction, due
to the elongational component of the shear flow \cite{abka08},
alternating strong shape fluctuations and relaxation into more circular shapes.
This mechanism occurs twice during every rotation cycle of the membrane,
as can also be appreciated in the supplementary movie files movie2.mp4 and
movie3.mp4. 
In order to characterize the UND motion, it may be useful to plot the asphericity as a function of the inclination angle, as reported in Fig.~\ref{fig:asphertilt_und}.
When $\phi=0.6$, the trajectories for the external vesicle are symmetric with respect to a vertical axis at a value
$\lla \theta \rra_{ext}^a > 0$. For clarity, we recall that in the TR case, the pattern of trajectories is similar but centered at $\lla \theta \rra_{ext}^a \simeq 0$. The internal vesicle undergoes SW with large variations of the 
inclination angle while the asphericity weakly varies. 
By increasing the filling fraction, the trajectories of the external vesicle
assume the characteristic shape of a closed ring, elongated along the vertical direction,
while the pattern of the internal vesicle is typical of TT motion.
The relative positions
of the centers of mass form closed fluctuating trajectories whose amplitudes decrease with
$\phi$, as illustrated in Fig.~\ref{fig:cm_und_lambda15_phi0608_gam20}
for the cases with
$\Delta_{ext}=1.23$, $\Delta_{int}=0.74$,
$\phi=0.6, 0.8$ at $(\Lambda, S)=(5.38, 138)$ of
Figs.~\ref{fig:und_lambda15_phi06_gam20} and \ref{fig:und_lambda15_phi08_gam20}.
\begin{figure}[h!]
\begin{center}
\includegraphics*[width=.35\textwidth,angle=0]{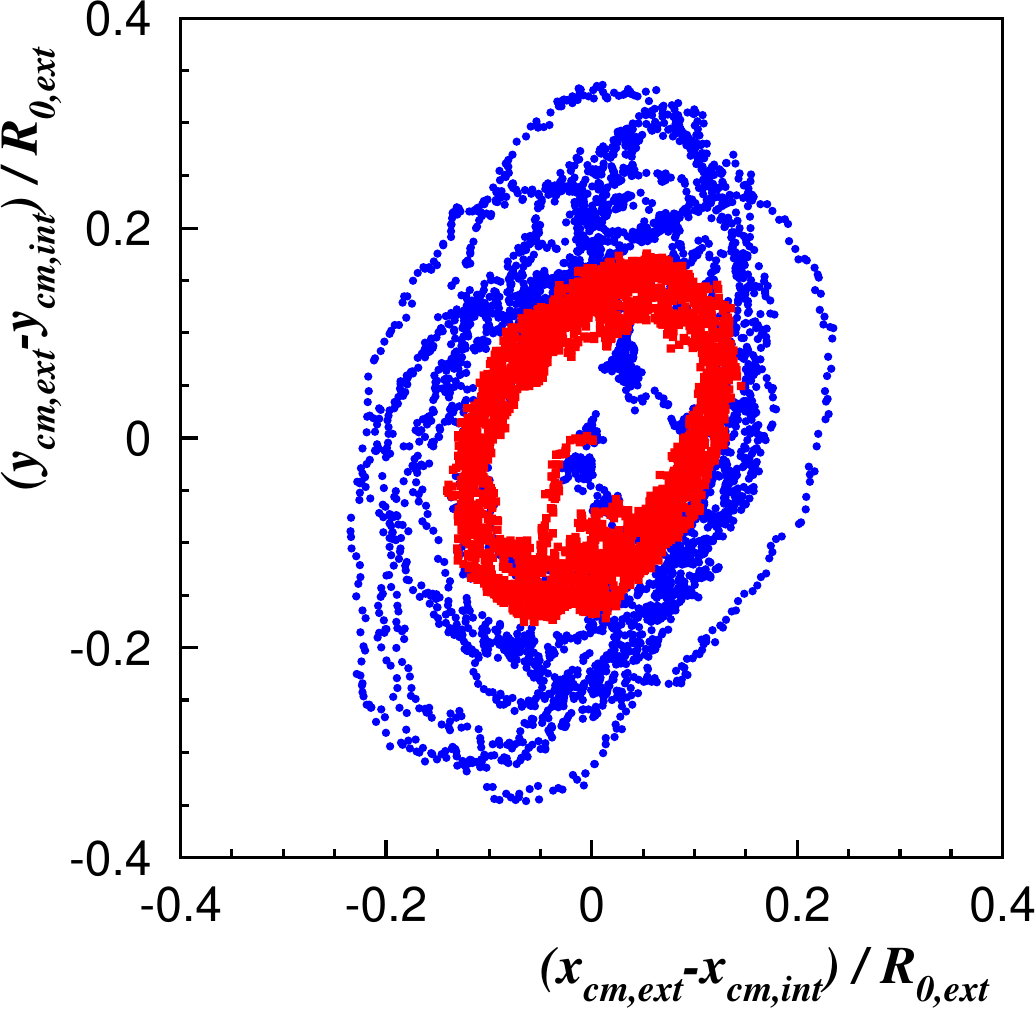}
\caption{Probability distribution of the center-of-mass relative positions
for a compound vesicle
corresponding to the cases shown  
in Fig.~\ref{fig:und_lambda15_phi06_gam20} (blue circles)
and in Fig.~\ref{fig:und_lambda15_phi08_gam20} (red squares).
\label{fig:cm_und_lambda15_phi0608_gam20}
}
\end{center}
\end{figure}
Lastly,
when the external and internal vesicle have the same swelling degree, 
we still observe UND motion for the external vesicle with the formation of more lobes. The motion of the internal vesicle is again controlled by the filling fraction resulting in
SW at $\phi=0.6$ and TT at $\phi=0.8$.
However, the amplitudes in the oscillations of $\theta_{ext}(t)$ and
$A_{ext}(t)$ are reduced with respect to the cases having $\Delta_{ext} > \Delta_{int}$.
Typical configurations of vesicles with $\Delta_{ext}=\Delta_{int}=0.74$, $\phi=0.6, 0.8$
at $(\Lambda, S)=(4.18, 229)$ 
are shown in Fig.~\ref{fig:conf_und} (c), (d), respectively.
The presence of UND motion with four rotating
lobes in the external vesicle has been also observed
in previous simulations with $\Delta_{ext}=\Delta_{int}$ at
$\phi=0.85$ and large shear rates \cite{kaou13}. In that case, however,
due to the lack of thermal fluctuations,
external and internal vesicles show radial symmetry, their centers are stationary,
and the internal vesicle performs SW motion, differently from the present work.
Finally, we add that the UND regime has never been observed in experiments, probably due to the difficulty of accessing high values of $\phi$ \cite{leva14}.

\section*{Conclusions}
We have studied the dynamics of a compound vesicle in a wall-bounded shear flow.
Encapsulating a vesicle into an external one adds extra degrees of freedom with respect
to the case of a single vesicle,
due to internal vesicle rotation, displacement, deformation, and to the hydrodynamic coupling between the internal and external membranes. 
Our outcomes have been obtained by using mesoscale hydrodynamic simulations of a two-dimensional model system, including thermal fluctuations. This latter feature, lacking in all previous
analytical and numerical approaches, has revealed to be fundamental in order
to observe characteristic trembling motion, as pointed out in experiments \cite{leva14}.
By varying the flow parameters $\Lambda$ and $S$, the filling fraction $\phi$, and
the internal excess length $\Delta_{int}$, the system is shown to manifest
a rich phenomenology in the dynamical behavior. At values $\Lambda \lesssim 1$, vesicles
exhibit tank-treading motion that is fully characterized. The external vesicle shows the same
inclination angle of a single vesicle at fixed $\Lambda$ and assumes a 
parallelogram-like shape, as quantified by the spectrum of shape amplitudes.
The internal vesicle shows a larger inclination angle whose difference with respect
to the external one does not depend on $\Lambda$ and $\phi$.
By increasing $\Lambda$ slightly beyond the value $\Lambda_c$ of the TT-TR transition, the 
external vesicle manifests trembling motion while the internal one swings. 
Higher values of $\Lambda$ trigger the transition to the tumbling motion of 
both vesicles.
At small filling fraction, the two vesicles appear desynchronized while larger
values of $\phi$ cause the mutual synchronization of vesicles. 
For large values of $\Lambda$, $S$, and $\phi$, the external vesicle shows undulating motion
\cite{kaou13}, characterized by periodic oscillations of inclination angle and 
asphericity while the membrane buckles.
This dynamic state exhibits unique patterns in the $A-\theta$ space that are shown to depend on the relative size and the swelling degree of both vesicles.
 The present numerical model, though limited to consider
a two-dimensional system, is able to provide quantitative agreement
with the experimental results of Ref.~\cite{leva14}. This is a remarkable feature that was previously shown to hold in describing correctly the flow patterns
of single vesicles in shear flow \cite{afik16}. In the present investigation, it was set 
$\eta_{in}=\eta_{out}$ and $\eta_{an} \geq \eta_{out}$ which implies that 
$\eta_{in} \leq \eta_{an}$. In the case of a single vesicle, this would correspond to a tank-treading condition. However, we find that the internal vesicle can be found in different dynamical states
due to the complex overall dynamics.
It would be interesting to consider which effects might be triggered
by setting $\eta_{in} > \eta_{an}$. Finally, we comment about the presence of boundary walls.
In our case, the degree of confinement, defined to be 
$\chi=2 R_{0,ext}/L_y$ \cite{kaou11}, is $0.35$. For this value of $\chi$, 
the study of single vesicles with the same numerical model \cite{lamu13,lamu15,lamu22} 
did not show relevant differences with respect to the analytical solutions of Keller and Skalak [14], corresponding to unbounded shear flow ($\chi=0$). 
We do not expect that for the considered value of $\chi$, walls
can quantitatively affect the present results with respect to unbounded shear
flows. Indeed, our results agree quantitatively with the ones of Ref.~\cite{leva14} whose experimental
setup is less confined than ours, being $\chi \simeq 0.25$.
It would be interesting to consider effects of strong confinement ($\chi > 0.4$) on the present phenomenology, as done for single vesicles \cite{kaou11}.

In conclusion, our model has allowed us to explore a large part of the space of controlling parameters so to both obtain results that are fully consistent with experiments, and
characterize the undulating motion \cite{kaou13}  providing a comprehensive study under different conditions. 
We hope that this latter result can stimulate further experimental investigations capable
to consider systems with high filling fractions at large values of $\Lambda$ and $S$.
Moreover, we believe that the present
approach can reveal very useful in order to address the dynamic behavior of compound vesicles in oscillatory shear flow \cite{nogu10} as well as of vesicles enclosing two or more vesicles, as in the case of vesosomes \cite{kisa04}.

%\section*{Author contributions}

\section*{Conflicts of interest}
There are no conflicts to declare.

\section*{Data availability}
Data are available from the author upon reasonable request.

\section*{Acknowledgements}
The work of AL was performed under the auspices of GNFM-INdAM.

%%%END OF MAIN TEXT%%%

%The \balance command can be used to balance the columns on the final page if desired. It should be placed anywhere within the first column of the last page.

\balance

%%%REFERENCES%%%

%\bibliography{biblio_ves} %.bib

\providecommand*{\mcitethebibliography}{\thebibliography}
\csname @ifundefined\endcsname{endmcitethebibliography}
{\let\endmcitethebibliography\endthebibliography}{}

\bibliographystyle{rsc} %the RSC's .bst file

\end{document}